\ifdefined\JSSMODE
  \documentclass[article]{jss}
\else
  \documentclass{article}
\fi

\usepackage{amsmath}
\usepackage{booktabs}
\usepackage{microtype}
\usepackage{listings}
\usepackage{xurl}
\usepackage{tikz}
\usetikzlibrary{arrows.meta,positioning,shapes.geometric,fit,calc}

\ifdefined\JSSMODE
\else
  \usepackage{lmodern}
  \usepackage{graphicx}
  \usepackage[margin=1in]{geometry}
  \usepackage[authoryear,round]{natbib}
  \usepackage{hyperref}
  \newcommand{\pkg}[1]{\texttt{#1}}
  
  \newcommand{\code}[1]{\texttt{#1}}
  \newcommand{\Plainauthor}[1]{}
  \newcommand{\Plaintitle}[1]{}
  \newcommand{\Shorttitle}[1]{}
  \newcommand{\Plainkeywords}[1]{}
  \newcommand{\Address}[1]{}
  \newcommand{\email}[1]{\texttt{#1}}
  \newcommand{\myabstract}{}
  \newcommand{\Abstract}[1]{\renewcommand{\myabstract}{#1}}
  \newcommand{\mykeywords}{}
  \newcommand{\Keywords}[1]{\renewcommand{\mykeywords}{#1}}
  \lstnewenvironment{CodeInput}{}{}
\fi

\newcommand{\PublicationDrawCount}{10,000}
\newcommand{\PublicationSmokingVop}{\$128,706 [\$0, \$279,088]}
\newcommand{\PublicationHousingVop}{\$30,853 [\$0, \$89,122]}
\newcommand{\PublicationSmokingDiscordance}{56.8\% (95\% Wilson interval 55.8--57.7\%)}
\newcommand{\PublicationHousingDiscordance}{73.3\% (95\% Wilson interval 72.4--74.1\%)}
\newcommand{\PublicationResultRows}{%
HPV Vaccination & -\$8,623 [-\$14,600, -\$4,901] & -\$14,133 [-\$26,567, -\$5,846] & \$0 [\$0, \$0] \\
Smoking Cessation & \$17,032 [-\$230,381, \$251,728] & -\$110,706 [-\$1,574,401, \$1,409,553] & \$128,706 [\$0, \$279,088] \\
Hepatitis C Therapy & -\$2,749 [-\$4,060, -\$1,938] & -\$12,257 [-\$18,945, -\$8,221] & \$0 [\$0, \$0] \\
Childhood Obesity Prevention & -\$5,178 [-\$13,739, -\$1,968] & -\$52,831 [-\$141,084, -\$19,679] & \$0 [\$0, \$0] \\
Housing Insulation & \$30,111 [-\$293,352, \$335,820] & -\$5,658 [-\$130,214, \$89,133] & \$30,853 [\$0, \$89,122] \\
}
\newcommand{\PublicationDecisionRows}{%
HPV Vaccination & Accept & Accept & 0.0\% [0.0\%, 0.0\%] & \$0 [\$0, \$0] \\
Smoking Cessation & Reject & Accept & 56.8\% [55.8\%, 57.7\%] & \$128,706 [\$0, \$279,088] \\
Hepatitis C Therapy & Accept & Accept & 0.0\% [0.0\%, 0.0\%] & \$0 [\$0, \$0] \\
Childhood Obesity Prevention & Accept & Accept & 0.0\% [0.0\%, 0.0\%] & \$0 [\$0, \$0] \\
Housing Insulation & Reject & Accept & 73.3\% [72.4\%, 74.1\%] & \$30,853 [\$0, \$89,122] \\
}

\newcommand{\SoftwareComparisonRows}{%
dampack & R & Manual & No & No \\
heemod & R & Manual & No & No \\
BCEA & R & Manual & No & No \\
hesim & R/C++ & Manual & Supported & No \\
DCEA & R & No & Native & No \\
dceasimR & R & No & Native & No \\
econeval & Python & Manual & No & No \\
heormodel & Python & Manual & No & No \\
PyHEOR & Python & Manual & No & No \\
rdecision & R & Manual & No & No \\
}

\author{Dylan A. Mordaunt\\Victoria University of Wellington}
\Plainauthor{Dylan A. Mordaunt}

\title{\pkg{vop\_poc\_nz}: A Python Framework for Distributional Cost-Effectiveness and Value of Perspective Analysis}
\Plaintitle{vop_poc_nz: A Python Framework for Distributional Cost-Effectiveness and Value of Perspective Analysis}
\Shorttitle{\pkg{vop\_poc\_nz}: Perspective and Distributional Analysis}

\Abstract{
  Health economic evaluations are sensitive to the choice of analytical perspective (e.g., health system vs. societal). 
  We present \pkg{vop\_poc\_nz}, a Python package for Distributional Cost-Effectiveness Analysis (DCEA), Markov cohort modeling, probabilistic sensitivity analysis, value of information, and explicit comparison of analytical perspectives.
  We define Value of Perspective (VoP) as a directional expected opportunity loss: the welfare loss, measured under a declared reference perspective, when decisions are selected under another perspective.
  Five synthetic Aotearoa New Zealand demonstration models produced deterministic decision discordance in two cases at a NZ\$20,000/QALY threshold. Seeded probabilistic analysis estimated mean per-person directional VoP of NZ\PublicationSmokingVop{} for smoking cessation and NZ\PublicationHousingVop{} for housing insulation (95\% simulation intervals in brackets); the other three cases had zero loss under the simulated strategy choices.
  These values test the software workflow and are not estimates for policy adoption.
  Release 0.2.3 supports Python 3.12--3.14 and records typed inputs, random seeds, software versions, and Arrow-schema identities for reproducible analysis.
}

\Plainkeywords{health economics, cost-effectiveness analysis, distributive cost-effectiveness analysis, economic evaluation, budget impact analysis, value of information, python, value of perspective}

\Address{
  Dylan A. Mordaunt\\
  Victoria University of Wellington\\
  Wellington, New Zealand\\
  E-mail: \email{dylan.mordaunt@vuw.ac.nz}
}

\begin{document}
\ifdefined\JSSMODE
\else
  \maketitle
  \begin{abstract}
  \myabstract
  \end{abstract}
\fi

\section{Introduction} \label{sec:intro}

Cost-effectiveness analysis (CEA) is widely used in health technology assessment to inform resource allocation \citep{drummond2015}.
The analytical \emph{perspective} determines which costs and benefits are included in CEA \citep{sanders2016}.
The two most common perspectives are the \emph{health system} perspective (focusing on direct medical costs) and the \emph{societal} perspective (including broader costs such as productivity losses and patient out-of-pocket expenses) \citep{drummond2015}.

Discrepancies between these perspectives can lead to \emph{decision discordance}, where an intervention is deemed cost-effective from one perspective but not the other \citep{claxton2010}.
For example, a vaccination program might be costly to the health system but generate significant productivity gains by preventing workforce absenteeism, making it cost-effective only from a societal perspective.
Guidelines from bodies like the International Decision Support Initiative (iDSI) and the World Health Organization (WHO) often recommend considering multiple perspectives \citep{wilkinson2016, tan-torres2003}. However, standard software tools typically treat perspective as a static parameter choice rather than a source of structural uncertainty to be quantified \citep{alarid2024, filipovic2017}.

We introduce \pkg{vop\_poc\_nz}, a Python package designed to:
\begin{enumerate}
  \item Implement a pipeline for CEA and Distributional CEA (DCEA) \citep{cookson2017}.
  \item Calculate the \emph{Value of Perspective} (VoP), a metric representing the opportunity cost of decision discordance.
  \item Provide Value of Information (VOI) analysis \citep{claxton2001}.
\end{enumerate}

The remainder of this paper is organized as follows. Section \ref{sec:methods} outlines the methodological framework, including the mathematical definitions of DCEA and VoP. Section \ref{sec:arch} describes the software architecture and design principles of \pkg{vop\_poc\_nz}. Section \ref{sec:comparison} compares the package with existing software alternatives. Section \ref{sec:cases} presents the case studies from Aotearoa New Zealand used to demonstrate the package's capabilities. Section \ref{sec:results} details the results of these analyses, highlighting the insights gained from the VoP metric. Finally, Section \ref{sec:discussion} discusses the implications for health policy and future software development.

\section{Methodological framework} \label{sec:methods}

The \pkg{vop\_poc\_nz} package implements Distributional Cost-Effectiveness Analysis (DCEA) and Value of Perspective (VoP) calculations alongside Markov cohort modeling.

\subsection{Cost-effectiveness analysis (CEA)}
Standard CEA compares the costs and health outcomes of two or more interventions to determine which represents the best value for money \citep{drummond2015}.
The primary metric is the Incremental Cost-Effectiveness Ratio (ICER), as defined in Equation~\ref{eq:icer} \citep{drummond2015}:
\begin{equation} \label{eq:icer}
ICER = \frac{\Delta C}{\Delta E} = \frac{C_{new} - C_{comparator}}{E_{new} - E_{comparator}}
\end{equation}
where $\Delta C$ and $\Delta E$ are the incremental costs and effects (e.g., QALYs), respectively.
An intervention is deemed cost-effective if its ICER is below a willingness-to-pay (WTP) threshold ($\lambda$) \citep{claxton2010}.
Alternatively, the Net Monetary Benefit (NMB) can be calculated (Equation~\ref{eq:nmb}) \citep{drummond2015}:
\begin{equation} \label{eq:nmb}
NMB = E \cdot \lambda - C
\end{equation}
A positive incremental NMB ($\Delta NMB > 0$) indicates cost-effectiveness.

\subsection{Markov modeling}
The core engine employs a time-homogeneous Markov cohort simulation \citep{briggs1998}. The model tracks a cohort of individuals through a set of mutually exclusive and collectively exhaustive health states (e.g., Healthy, Sick, Dead) over discrete time cycles (typically annual).
In each cycle $t$, the proportion of the cohort in state $j$, $P_j(t)$, is updated based on transition probabilities derived from the transition matrix $M$ (Equation~\ref{eq:markov}) \citep{briggs1998}:
\begin{equation} \label{eq:markov}
P(t+1) = P(t) \times M
\end{equation}
Costs ($C$) and health outcomes ($Q$, typically QALYs) are accumulated for each state. To support multi-perspective analysis, the model simultaneously tracks two sets of cost vectors: $C_{HS}$ (Health System) and $C_{Soc}$ (Societal), where $C_{Soc} = C_{HS} + C_{Prod} + C_{OOP}$ (Productivity and Out-of-Pocket costs).

\subsection{Distributional cost-effectiveness analysis (DCEA)}
Standard CEA maximizes total population health (e.g., QALYs) subject to a budget constraint. DCEA extends this by incorporating equity concerns, explicitly trading off total health maximization against the reduction of health inequality \citep{cookson2017}.

Let $\pi_g$ denote subgroup $g$'s population share, with
$\sum_g\pi_g=1$. A standard Atkinson social welfare calculation evaluates
the complete distribution of health using equally distributed equivalent
(EDE) health:
\begin{equation} \label{eq:nmb_eq}
H_{EDE}=\left(\sum_g \pi_g H_g^{1-\epsilon}\right)^{1/(1-\epsilon)},
\quad \epsilon\ne1,
\end{equation}
where $H_g$ is subgroup lifetime health and $\epsilon$ is the inequality
aversion parameter \citep{atkinson1970,asaria2016}. Strategies may then be
compared using changes in EDE health after incorporating the subgroup
distribution of health opportunity costs \citep{cookson2017,lovekoh2019}.

For software demonstrations that supply subgroup incremental outcomes but not
a complete lifetime-health distribution, the package also exposes a
transparent weighted-NMB approximation. The unnormalised level-dependent
weight is
\begin{equation} \label{eq:weights}
w_g = \left( \frac{H_{ref}}{H_g} \right)^\epsilon
\end{equation}
and the population-normalised weight is
\begin{equation} \label{eq:normalised_weights}
\widetilde w_g=\frac{w_g}{\sum_h\pi_h w_h}.
\end{equation}
The corresponding approximation is
\begin{equation} \label{eq:weighted_nmb}
NMB_{eq}=\sum_g\pi_g\widetilde w_g(\lambda\Delta Q_g-\Delta C_g).
\end{equation}
When $\epsilon=0$, $\widetilde w_g=1$ and Equation~\ref{eq:weighted_nmb}
reduces to population-weighted conventional NMB. This approximation is not a
substitute for full distributional modeling. Its results depend on the
declared reference-health measure, subgroup definitions, population shares,
and treatment of opportunity costs.

\subsection{Value of information (VOI)}
Decision uncertainty arises from parameter uncertainty. VOI analysis quantifies the value of reducing this uncertainty through further research \citep{claxton2001}.
\begin{itemize}
    \item \textbf{Expected Value of Perfect Information (EVPI)}: The expected value of removing all parameter uncertainty. It is calculated as the difference between the expected value of the decision made with perfect information and the decision made with current information. The package also supports calculating \textbf{Population EVPI} by scaling this per-person value by the affected population size.
    \item \textbf{Expected Value of Partial Perfect Information (EVPPI)}: The expected value of removing uncertainty for a specific subset of parameters.
\end{itemize}

\subsection{Value of perspective (VoP)}
The choice of analytical perspective can alter the selected strategy
\citep{mordaunt2024ssrn}. We define VoP as a directional loss relative to an
explicitly declared reference perspective. It does not estimate which
perspective is normatively correct.

Let $r$ denote the reference perspective and $p$ the perspective used to make
the decision. For parameters $\theta$, define
$d_p(\theta)=\arg\max_d NMB_p(d,\theta)$. The deterministic
perspective-induced decision loss is
\begin{equation} \label{eq:discordance}
L_{p\rightarrow r}(\theta)=
\max_d NMB_r(d,\theta)-NMB_r(d_p(\theta),\theta).
\end{equation}
The loss is zero when the perspective-$p$ decision also maximises NMB under
the reference perspective. Otherwise, it measures forgone net benefit under
the declared reference welfare standard.

The probabilistic VoP is the expectation over the joint parameter distribution:
\begin{equation} \label{eq:evop}
VoP_{p\rightarrow r}=E_{\theta}\left[L_{p\rightarrow r}(\theta)\right].
\end{equation}
The demonstrations use $p=HS$ and $r=Soc$, so their values should be read as
$VoP_{HS\rightarrow Soc}$. Reversing the reference perspective can produce a
different result. When decision makers do not accept a single reference
perspective, the package reports the directional matrix of losses rather than
averaging across unelicited normative probabilities.

\subsection{Budget impact analysis (BIA)}
Budget Impact Analysis estimates the financial consequences of adopting a new intervention for a specific budget holder over a short-to-medium term horizon (typically 1--5 years) \citep{sullivan2014}.
Unlike CEA, which focuses on efficiency (value for money), BIA focuses on affordability \citep{sullivan2014}.
We calculate the net budget impact ($BI$) as shown in Equation~\ref{eq:bia}:
\begin{equation} \label{eq:bia}
BI = \sum_{t=1}^{T} (C_{New}(t) - C_{Comparator}(t)) \cdot P_{uptake} \cdot N_{pop}
\end{equation}
where $N_{pop}$ is the size of the eligible population and $P_{uptake}$ is the uptake rate of the new intervention.

\subsection{Normative framework}
The use of DCEA and VoP implies specific normative commitments that must be transparently communicated.
\begin{itemize}
    \item \textbf{Inequality Aversion}: The parameter $\epsilon$ in Equation~\ref{eq:weights} is not an empirical constant but a moral choice reflecting society's tolerance for health inequality \citep{atkinson1970}. A value of $\epsilon=10.9$ (often cited in literature) implies a strong priority for the worst-off, while $\epsilon=0$ implies no special priority. Our default is $\epsilon=0.5$, representing mild aversion, following the illustrative examples in \cite{cookson2017}.
    \item \textbf{Reference perspective}: VoP requires a declared reference welfare standard. The demonstrations use the societal perspective, but the software retains directionality and can report the reverse comparison. Treating societal NMB as the reference remains a utilitarian value judgment rather than an empirical finding \citep{brouwer2008}.
    \item \textbf{Opportunity Costs}: Both CEA and DCEA assume that resources used for an intervention displace other health-generating activities. The WTP threshold $\lambda$ represents the shadow price of this displacement \citep{claxton2010}.
\end{itemize}
Users must explicitly acknowledge these value judgments when interpreting results.

\section{Software design and implementation} \label{sec:arch}

Figure~\ref{fig:contract-architecture} summarises the current contract-first
architecture. The diagram is authored in PGF/TikZ so that the paper's
architecture remains deterministic, editable, and typography-consistent with
the surrounding text.
\begin{figure}[htbp]
\centering
\resizebox{\textwidth}{!}{%
\begin{tikzpicture}[node distance=8mm and 12mm, >=Latex,
  box/.style={draw, rounded corners, align=center, minimum width=28mm, minimum height=9mm, fill=blue!6},
  gate/.style={draw, diamond, aspect=2, align=center, fill=orange!12}]
  \node[box] (inputs) {Parameters\\and evidence};
  \node[box, right=of inputs] (contracts) {Pydantic v2\\method contracts};
  \node[box, right=of contracts] (kernel) {Calculation\\kernels};
  \node[box, below=of contracts] (arrow) {Arrow/JSON\\result envelope};
  \node[gate, below=of kernel] (gate) {Validation\\gates};
  \node[box, right=of gate] (reports) {Reports, figures\\and paper receipts};
  \draw[->] (inputs) -- (contracts);
  \draw[->] (contracts) -- (kernel);
  \draw[->] (kernel) -- (gate);
  \draw[->] (gate) -- (reports);
  \draw[->] (kernel) |- (arrow);
\draw[->] (arrow) -| (gate);
\end{tikzpicture}
}
\caption{Contract-first VOP architecture and publication evidence path.}
\label{fig:contract-architecture}
\end{figure}

The \pkg{vop\_poc\_nz} package combines typed domain models with functional
calculation kernels. It uses \pkg{numpy}, \pkg{pandas}, \pkg{scipy}, and
\pkg{matplotlib}; Pydantic v2 validates method and result contracts; and
\pkg{pandera} validates DataFrame boundaries. Apache Arrow schemas, IPC files,
and Parquet provide language-neutral interchange with schema fingerprints.
Optional Polars and experimental backends are kept behind explicit dependency
features so that they cannot silently change the reference calculation.

\subsection{Core components}
The architecture separates the mathematical core from the analysis pipeline and reporting layer, ensuring a clear abstraction between model inputs and computational logic.
\begin{itemize}
  \item \code{cea\_model\_core}: The computational engine. It implements vectorized Markov cohort models, allowing for rapid execution of thousands of probabilistic iterations. The state distribution is computed over a time horizon $T$ using matrix multiplication, where transition matrices are 3D tensors (simulations $\times$ states $\times$ states). It supports multiple simultaneous perspectives by tracking costs and utilities in parallel accumulators.
  \item \code{dcea\_equity\_analysis}: A specialized module for equity calculations. It takes subgroup-disaggregated results and applies the Atkinson social welfare function to compute equity-weighted metrics.
  \item \code{value\_of\_information}: Implements the VoP and VOI logic. It uses Monte Carlo simulation results to estimate the expected opportunity loss of decision uncertainty.
\end{itemize}

\subsection{Configuration and abstraction}
All model parameters are defined in external \code{YAML} configuration files, decoupling the data from the code.
This allows researchers to modify transition probabilities, costs, and utility values without altering the underlying Python codebase.
The \code{ModelParameters} class validates these inputs before they are passed to the computational engine.

\subsection{Design patterns}
The high-level workflow uses a functional pipeline with the following stages:
\begin{enumerate}
    \item \textbf{Ingestion}: Parameters are loaded from YAML configuration files.
    \item \textbf{Simulation}: The \code{MarkovModel} class executes the cohort simulation for each intervention and perspective.
    \item \textbf{Aggregation}: Results are aggregated by subgroup and perspective.
    \item \textbf{Analysis}: PSA, DCEA, and VoP analyses are performed on the aggregated data.
    \item \textbf{Reporting}: The \code{reporting} module generates static plots and Markdown reports.
\end{enumerate}
This separation permits post-processing changes that do not alter simulated
state trajectories. Changes to parameters that affect transitions, costs,
utilities, subgroup opportunity costs, or stochastic draws require a new run;
the evidence manifest records that dependency boundary.

\section{Comparison with existing software} \label{sec:comparison}

We performed a systematic software review on 21 July 2026 using seven
pre-specified GitHub repository-search queries covering CEA, Markov modelling,
PSA/VOI, DCEA, and ``value of perspective.'' The search retrieved 202 records;
after removing 27 duplicates, 175 unique repositories were screened, 17 were
assessed in full text, and 10 reusable open-source packages were included.
Eligibility criteria, exact search strings, flow counts, exclusions, versions,
feature evidence, and limitations are recorded in the machine-readable review
ledger and its protocol. Public README files, tagged documentation, package
websites, and release metadata were inspected. A ``No'' in
Table~\ref{tab:software_comparison} means that no named feature was located in
those materials, not that an analyst could not implement the calculation using
the package.

\begin{table}[ht]
\centering
\resizebox{\textwidth}{!}{%
\begin{tabular}{lcccc}
\hline
Package & Language & Multi-perspective & DCEA & Directional VoP \\
\hline
\SoftwareComparisonRows
\textbf{\pkg{vop\_poc\_nz}} & \textbf{Python} & \textbf{Native} & \textbf{Native} & \textbf{Native} \\
\hline
\end{tabular}
}
\caption{Systematically screened open-source health-economic software. Versions and evidence links are recorded in the review ledger; ``Manual'' denotes separate analyst-configured runs.}
\label{tab:software_comparison}
\end{table}

\subsection{Contribution}
\pkg{vop\_poc\_nz} is native to the Python ecosystem and exposes perspective
comparison as a typed workflow. Within the 10 included packages and versions,
it was the only package whose public API named and tested a directional VoP
calculation. This bounded finding is not a claim about all software, unpublished
code, or calculations that users could construct from lower-level primitives.

\section{Case studies} \label{sec:cases}

To demonstrate the capabilities of \pkg{vop\_poc\_nz}, we applied the framework to five synthetic interventions framed in an Aotearoa New Zealand context: HPV Vaccination, Smoking Cessation, Hepatitis C Therapy, Childhood Obesity Prevention, and Housing Insulation.
The cases are software fixtures, not health technology assessments. Their inputs combine illustrative values with selected published sources and have not undergone the evidence synthesis, clinical validation, M\={a}ori governance, or stakeholder review required for policy use. Numerical results therefore demonstrate computation and reporting behavior only.
These cases were selected to represent a spectrum of intervention types, ranging from traditional clinical technologies (HPV Vaccination) to broader public health and social policy measures (Housing Insulation).
This diversity allows us to test the sensitivity of the VoP metric to different cost structures and societal spillover effects.
Specifically, we contrast a vaccine with high health system costs and clear health benefits against a housing intervention where the primary costs are non-health (retrofitting) but the benefits span health, energy savings, and productivity.

\subsection{Model structure}
Each intervention was modeled using a Markov cohort simulation with annual cycles over a lifetime horizon.
Parameters were assembled for demonstration from selected literature and illustrative adjustments (see Supplement C). They should not be interpreted as current New Zealand parameter estimates.
We evaluated two perspectives:
\begin{enumerate}
    \item \textbf{Health System}: Direct medical costs and health gains (QALYs).
    \item \textbf{Societal}: Adds productivity gains and broader societal costs. The framework supports both the \emph{Human Capital} method (valuing all lost time) and the \emph{Friction Cost} method (valuing time until replacement), with the Human Capital method used as the default for these case studies.
\end{enumerate}
Costs and health outcomes were discounted at 3\% per annum as an illustrative setting. Analysts preparing a live assessment should use the current Te P\={a}taka Whaioranga Pharmac methods guidance and test alternative discount rates \citep{pharmac2020}.

\subsection{Evaluation framework}
The evaluation proceeds in the following stages:
\begin{enumerate}
    \item \textbf{Deterministic Analysis}: We calculate the Incremental Cost-Effectiveness Ratio (ICER) for each perspective using base-case parameter values. This step identifies potential decision discordance.
    \item \textbf{Deterministic Sensitivity Analysis (DSA)}: We perform one-way sensitivity analyses to identify which parameters (e.g., discount rates, utility weights) most influence the results, visualized via Tornado diagrams.
    \item \textbf{Probabilistic Sensitivity Analysis (PSA)}: We run \PublicationDrawCount{} seeded Monte Carlo draws for the demonstration. Mean-one lognormal arm-level multipliers use coefficients of variation of 10\% for health-system costs, 15\% for societal costs, and 5\% for QALYs; QALY draws are shared across perspectives. This synthetic uncertainty model tests the workflow and is not an empirically estimated joint distribution.
    \item \textbf{Cost of Illness (COI)}: Prior to the full CEA, we estimate the total economic burden of the disease under the status quo, disaggregated by payer (Health System vs. Societal).
    \item \textbf{Budget Impact Analysis (BIA)}: We estimate the aggregate financial consequences of adopting the intervention over a 5-year horizon, considering the eligible population size.
    \item \textbf{Distributional and VoP Analysis}: We apply the Atkinson inequality aversion parameter ($\epsilon=0.5$) to estimate the Equity-Weighted Net Monetary Benefit ($NMB_{eq}$). Concurrently, we calculate the Value of Perspective (VoP) to quantify the opportunity cost of ignoring societal costs.
\end{enumerate}

\subsection{Monte Carlo assurance and inferential scope}
The software records the random seed, draw count, package version, input
digest, and Arrow schema identity with each typed result. Diagnostic functions
calculate running means, Monte Carlo standard errors, and simulation intervals
for perspective-loss estimates. A policy analysis should pre-specify a
precision target, increase draws until the Monte Carlo standard error is small
relative to the decision margin, and examine whether strategy-selection and
tail estimates are stable. The present results use \PublicationDrawCount{} draws, equal-tailed
95\% simulation intervals for continuous outcomes, Wilson intervals for
decision probabilities, and Monte Carlo standard errors recorded in the result
dataset. These intervals characterize the declared synthetic uncertainty model;
they do not represent empirical parameter uncertainty.

\subsection{Roadmap of results}
\begin{figure}[ht]
\centering
\includegraphics[width=0.8\textwidth]{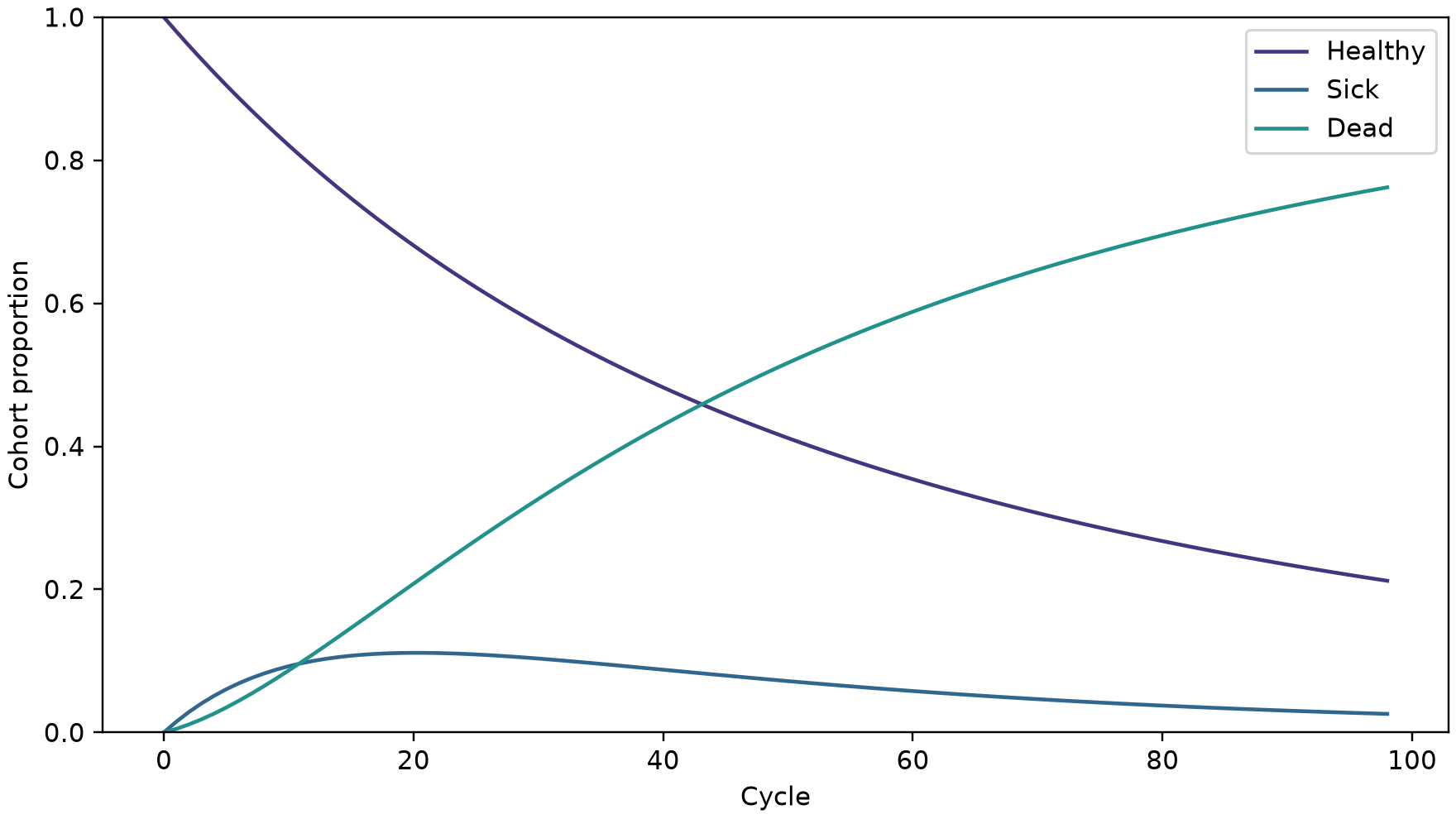}
\caption{Markov trace for the synthetic HPV vaccination model (new treatment). The plot shows the proportion of the cohort in each health state over time.}
\label{fig:markov_trace}
\end{figure}

The results section presents the following outputs:
\begin{itemize}
    \item \textbf{Model Structure}: Figure~\ref{fig:markov_trace} shows the generated cohort-state dynamics.
    \item \textbf{Cost-Effectiveness}: Table~\ref{tab:results} and Figure~\ref{fig:ce_plane} (Cost-Effectiveness Plane) compare the efficiency outcomes across perspectives.
    \item \textbf{Perspective Uncertainty}: Figure~\ref{fig:violin} (Delta NMB Violin Plot) and Figure~\ref{fig:ceac} (CEAC) visualize the impact of perspective choice on decision confidence.
    \item \textbf{Uncertainty}: Figure~\ref{fig:ceac} reports cost-effectiveness probabilities with Wilson 95\% intervals.
\end{itemize}

Figure~\ref{fig:markov_trace} shows the resulting HPV Vaccination Markov trace
for the new-treatment arm.

\subsection{Results} \label{sec:results}
Table~\ref{tab:results} reports median simulated incremental cost-effectiveness
ratios (ICERs) and mean directional VoP with equal-tailed 95\% simulation
intervals. Ratio intervals are wide when incremental QALYs approach zero and
should be interpreted together with NMB and decision probabilities.

\begin{table}[ht]
\centering
\resizebox{\textwidth}{!}{%
\begin{tabular}{lrrr}
\hline
Intervention & Health System ICER & Societal ICER & $VoP_{HS\rightarrow Soc}$ \\
\hline
\PublicationResultRows
\hline
\end{tabular}
}
\caption{Regenerated synthetic results. ICER entries are medians and directional VoP entries are means; brackets are equal-tailed 95\% simulation intervals (NZ dollars per person).}
\label{tab:results}
\end{table}

To explicitly quantify the value of adopting the societal perspective, Table~\ref{tab:vop_analysis} presents the Value of Perspective (VoP) analysis.
This table highlights the decision discordance -- where the optimal strategy differs between perspectives -- and the resulting opportunity cost (VoP) of adhering to the narrower health system perspective.

\begin{table}[ht]
\centering
\resizebox{\textwidth}{!}{%
\begin{tabular}{lllrr}
\hline
Intervention & Health-system & Societal & Discordance probability & $VoP_{HS\rightarrow Soc}$ \\
\hline
\PublicationDecisionRows
\hline
\multicolumn{5}{l}{\footnotesize \textit{Note: Decisions and probabilities use a WTP threshold of \$20,000/QALY.}} \\
\end{tabular}
}
\caption{Directional VoP analysis for synthetic demonstrations. VoP is the per-person loss under societal NMB when the strategy is selected using health-system NMB; it does not establish that the societal perspective is normatively correct.}
\label{tab:vop_analysis}
\end{table}

\begin{figure}[ht]
\centering
\includegraphics[width=1.0\textwidth]{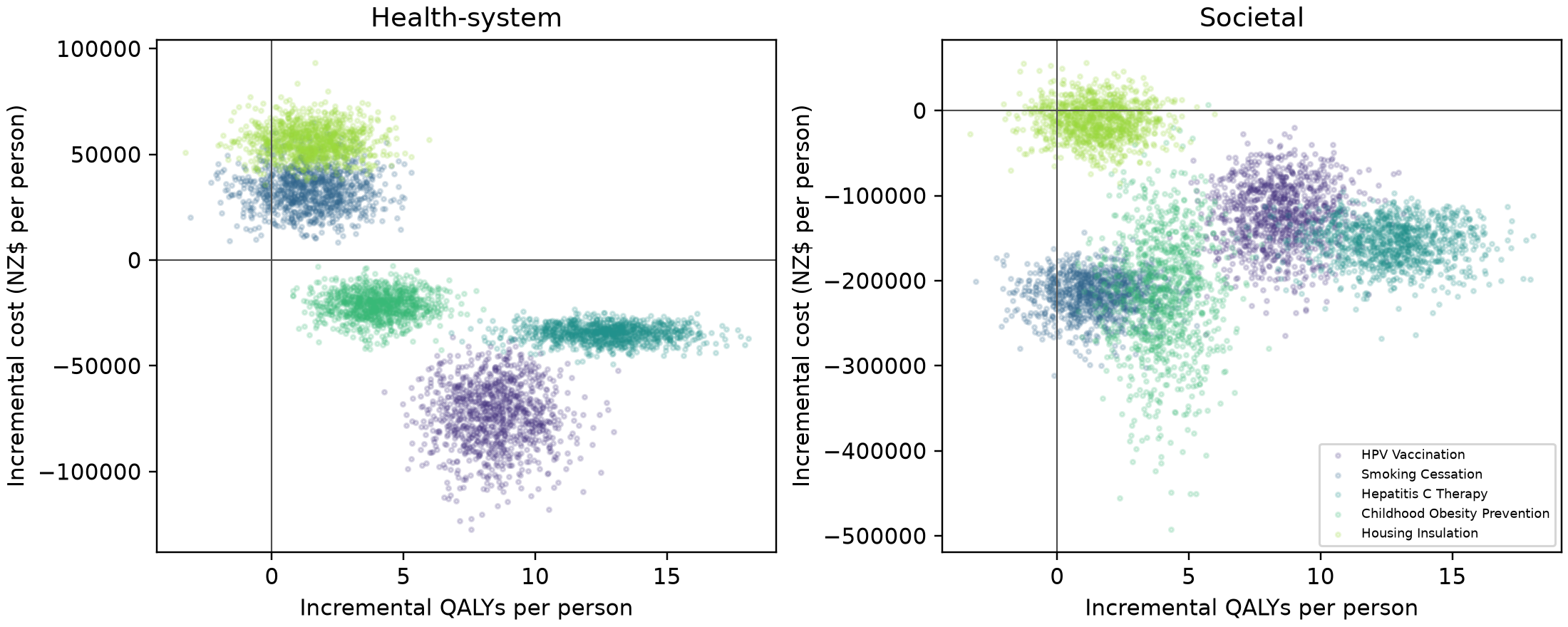}
\caption{Comparative cost-effectiveness planes from the manifest-backed draws. The left panel shows the health-system perspective and the right panel the societal perspective. Points in the bottom-right quadrant indicate lower modeled net costs and higher modeled health benefits.}
\label{fig:ce_plane}
\end{figure}

Figure~\ref{fig:ce_plane} visualizes the cost-effectiveness plane, highlighting the shift in the cost-effectiveness cloud when moving from the health system to the societal perspective.

\subsection{Value of perspective}
\begin{figure}[ht]
\centering
\includegraphics[width=1.0\textwidth]{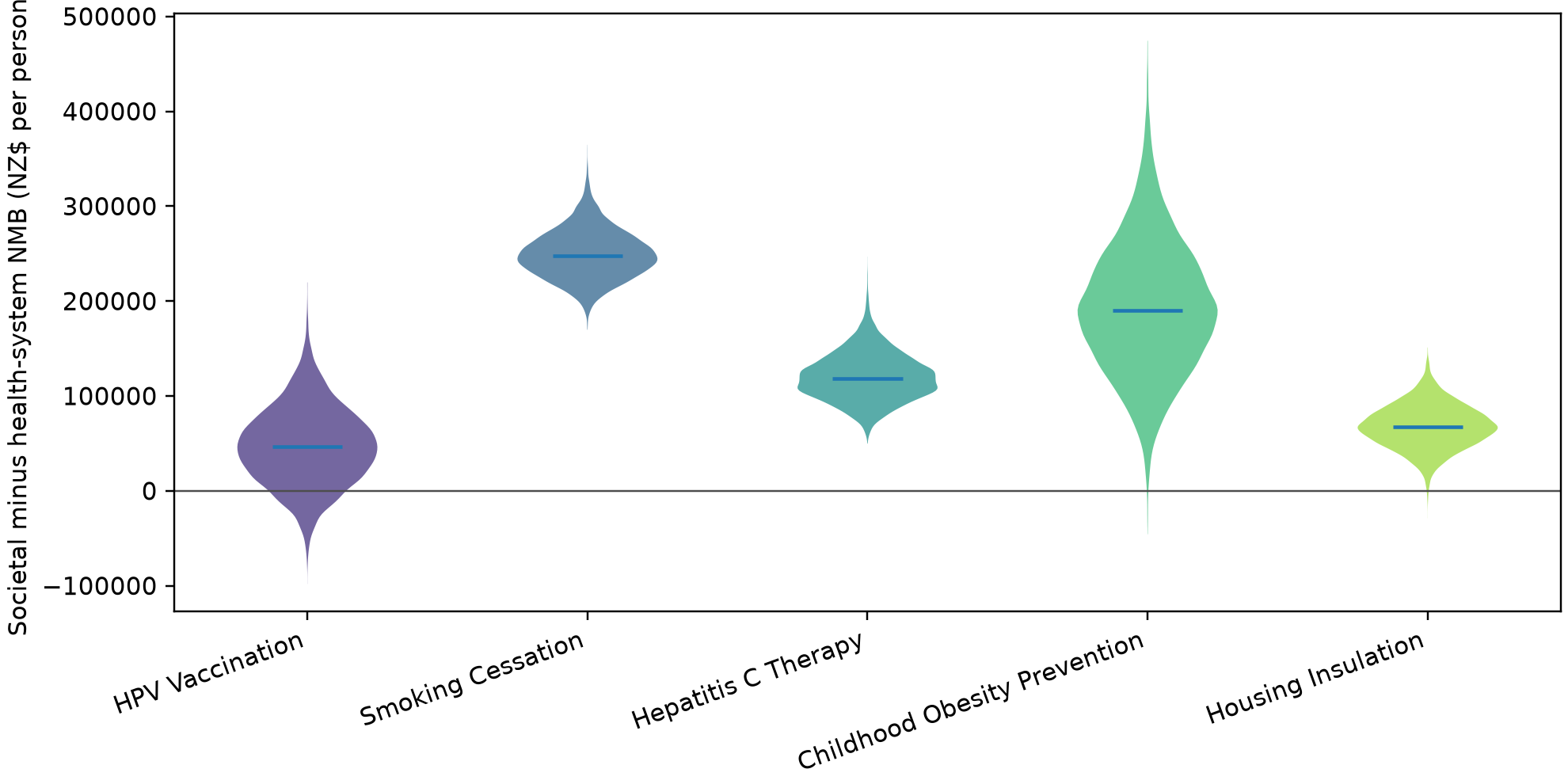}
\caption{Violin plot of delta net monetary benefit ($NMB_{soc} - NMB_{HS}$) at a NZ\$50,000/QALY threshold. Distributions shifted to the right of zero indicate higher modeled NMB under the societal perspective. Width represents the estimated density of the difference.}
\label{fig:violin}
\end{figure}

Within the deterministic synthetic inputs, all interventions were cost-effective under both perspectives at the NZ\$50,000 threshold, while two of five produced decision discordance at NZ\$20,000/QALY. Under the declared uncertainty model, discordance probabilities were \PublicationSmokingDiscordance{} for smoking cessation and \PublicationHousingDiscordance{} for housing insulation.
For example, at the lower threshold, Housing Insulation was rejected under the health-system perspective but accepted under the societal perspective, generating a positive $VoP_{HS\rightarrow Soc}$.
Figure~\ref{fig:violin} illustrates the distribution of the Delta Net Monetary Benefit ($NMB_{soc} - NMB_{HS}$) for each intervention.
Most synthetic distributions are shifted to the right of zero, although the magnitude and variance of the perspective difference vary.
Smoking Cessation shows a long right tail in the demonstration draws. This pattern is consistent with uncertain productivity gains omitted from the health-system perspective, but its magnitude should not be interpreted without additional draws and empirically validated inputs.

\subsection{Uncertainty analysis}
\begin{figure}[ht]
\centering
\includegraphics[width=0.8\textwidth]{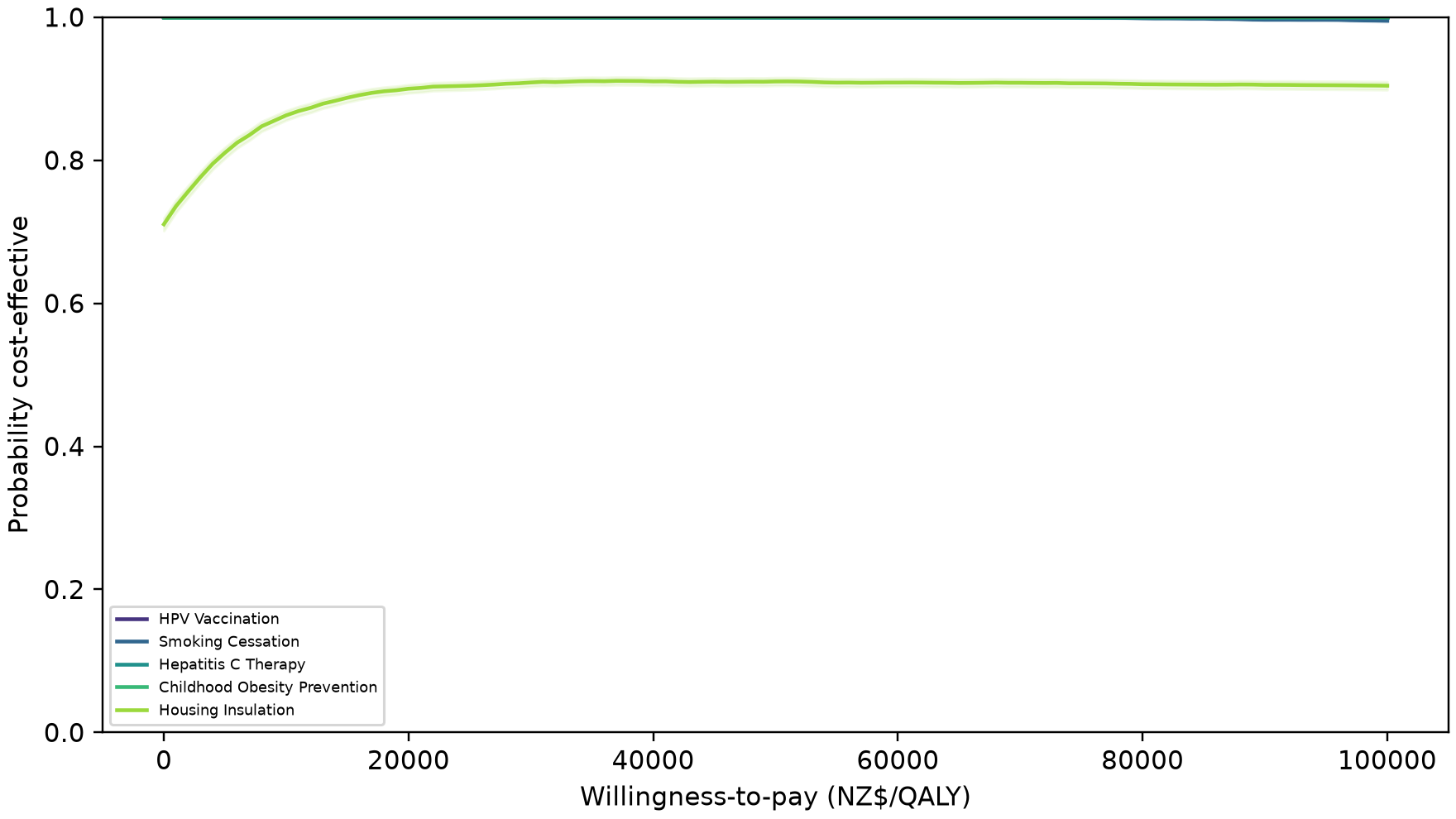}
\caption{Cost-effectiveness acceptability curve under the societal perspective. The plot shows the probability that each synthetic intervention is cost-effective across willingness-to-pay thresholds.}
\label{fig:ceac}
\end{figure}

Figure~\ref{fig:ceac} presents the Cost-Effectiveness Acceptability Curve (CEAC), showing the probability that each intervention is cost-effective at varying willingness-to-pay thresholds. Shaded bands are Wilson 95\% intervals for the finite Monte Carlo sample.

\section{Usage and reproducibility} \label{sec:usage}

The package provides both a command-line interface (CLI) for standard workflows and a Python API for custom analyses.

\subsection{Command-line interface}
The CLI is the primary entry point for running the full analysis pipeline.
\begin{CodeInput}
# Initialize a new project with default configuration
vop-poc-nz init my_project
cd my_project

# Run the full pipeline (CEA, DCEA, VoP)
vop-poc-nz run --output-dir results
\end{CodeInput}

\subsection{Python API}
For more granular control, users can import core classes directly.
\begin{CodeInput}
from vop_poc_nz.cea_model_core import MarkovModel
from vop_poc_nz.dcea_equity_analysis import calculate_equity_metrics

# Initialize model
model = MarkovModel(params, perspective="societal")
results = model.run()
\end{CodeInput}

\subsection{Reproducing case studies}

The reproducibility target is release \pkg{vop\_poc\_nz} 0.2.3 and the exact
source revision recorded by the result manifest. Python 3.12--3.14 is
supported. The frozen \code{uv.lock} and \code{pixi.lock} files define the
environment; Pydantic and Arrow contracts record typed inputs and output
schema identities. The reference verification commands are:
\begin{CodeInput}
pixi run verify
uv run vop-poc-nz run --output-dir output
uv run python scripts/publication_gate.py
\end{CodeInput}
The CLI accepts an explicit parameter file through \code{--parameters}. A
result intended for comparison must include the input digest, random seed,
draw count, package version, source revision, and artifact hashes. Sourceright
is pinned for source-provenance checks and AuthenText for editorial review;
both run as external tools and are not embedded in the package.
The repository examples and method contracts demonstrate how to:
\begin{itemize}
    \item validate perspective and equity parameters before execution;
    \item run seeded probabilistic and perspective diagnostics;
    \item serialize typed results to Arrow or Parquet; and
    \item generate figures and evidence manifests from recorded outputs.
\end{itemize}
The analytical tables and plotted numerical results in this article are
regenerated by \path{scripts/regenerate_manuscript_results.py}. The result
manifest binds the input and generator hashes, source revision, seed, draw
count, environment versions, uncertainty specification, and every generated
artifact hash.

\section{Discussion} \label{sec:discussion}

The \pkg{vop\_poc\_nz} package demonstrates the feasibility of operationalizing the Value of Perspective and DCEA within a unified computational framework.
The demonstrations show that analytical perspective can change a modeled decision when omitted cost categories are large relative to the decision margin.
By making normative assumptions explicit and quantifying the decision uncertainty associated with perspective choice, the tool enhances the transparency of health economic evaluations.
The directional loss metric provides one way to report the modeled opportunity cost of using a narrower perspective, conditional on the reference welfare standard.
This section discusses the implications of these findings for decision-making, the normative trade-offs inherent in equity-weighted analysis, and the computational considerations for scaling this approach.

\subsection{Perspective difference and decision discordance}
In the synthetic housing-insulation and smoking-cessation models, non-health costs and effects were large enough to alter NMB and, at some thresholds, the selected strategy.
The difference visualized in Figure~\ref{fig:violin} represents the modeled change in NMB when the analytical boundary is widened.
Quantifying this difference shows what is lost under the declared societal reference perspective when the health-system decision is used.
For example, the long right tail for smoking cessation reflects draws with larger productivity gains. Additional simulation and empirical validation would be needed to determine whether that pattern supports a policy conclusion.

\subsection{Normative and governance implications}
The integration of DCEA highlights that efficiency and equity are often competing objectives.
The parameter $\epsilon$ (inequality aversion) acts as a ``tuning knob'' for social justice.
Allowing users to vary $\epsilon$ supports sensitivity analysis of a declared value judgment; it does not convert the ethical choice itself into an empirical question.
Users must be transparent about these choices; a high $\epsilon$ value implies a willingness to sacrifice total population health to reduce inequality, a trade-off that requires democratic legitimacy.

Analyses involving M\={a}ori data require governance beyond technical validation.
Te Mana Raraunga defines M\={a}ori data sovereignty in terms of M\={a}ori rights and
interests in the collection, ownership, governance, and application of M\={a}ori
data \citep{temanararaunga2018}. Subgroup definitions, baseline-health
estimates, equity weights, access controls, and publication decisions should
therefore be developed with appropriate M\={a}ori governance and engagement and
in a manner consistent with te Tiriti o Waitangi. The present synthetic cases
used no individual-level M\={a}ori data and received no M\={a}ori stakeholder or
governance review. They cannot establish M\={a}ori equity effects or support
resource-allocation recommendations.

\subsection{Limitations}
The paper evaluates software behavior using synthetic models rather than
validated health technology assessments. Parameter selection was not based on
a systematic evidence review, model calibration, or external clinical
validation. The reported intervals propagate a declared synthetic uncertainty
model and do not repair those evidence limitations. The results do not
support conclusions about the effectiveness, cost effectiveness, budget
impact, or equity effects of the named interventions in Aotearoa New Zealand.

The directional VoP metric is conditional on the selected reference
perspective. It measures neither social consensus nor uncertainty about the
morally correct perspective. The weighted-NMB approximation also does not
replace full DCEA based on distributions of lifetime health and health
opportunity costs. Results can change with subgroup definitions, population
shares, baseline-health measures, inequality-aversion values, and the
reference perspective.

The finite Monte Carlo sample constrains precision. The generated dataset
therefore reports Monte Carlo standard errors in addition to the displayed
simulation intervals. A policy application should pre-specify precision and
convergence criteria and increase the draw count until they are met rather than
adopting the demonstration count as a universal minimum.

Current releases use uv and Pixi lockfiles, structured logging, dynamic
versioning, and CI across Python 3.12--3.14.
The repository includes Scalene profiling and experimental backend lanes.
Parallel or alternative backends must demonstrate numerical equivalence to
the reference kernels before promotion.

\subsection{Policy brief generation}
The ultimate goal of health economic modeling is to inform decision-making.
To this end, the package includes a prototype policy brief generator that translates the complex outputs of DCEA and VoP analysis into plain-language summaries.
An example of this output is provided in the supplementary materials.
These briefs highlight the key drivers of decision discordance and the equity implications of funding choices, bridging the gap between technical analysts and political decision-makers.
Future work will refine this module to support customizable templates for different stakeholder groups.

\section{Summary} \label{sec:summary}

\pkg{vop\_poc\_nz} provides a Python framework for explicit comparison of analytical perspectives and distributional outcomes.
Its directional VoP metric reports the modeled loss from selecting a strategy under one perspective and evaluating it under another declared reference perspective.
The integration of Distributional CEA further ensures that equity trade-offs are transparently weighed against efficiency gains.
We hope this open-source tool will encourage the broader adoption of multi-perspective and equity-informative modeling in health policy.

\section*{Declarations}

\subsection*{Funding}
The authors declare that no funds, grants, or other support were received during the preparation of this manuscript.

\subsection*{Conflicts of interest}
The authors have no relevant financial or non-financial interests to disclose.

\subsection*{Ethics}
This software study used synthetic demonstration models and no human
participants, identifiable records, or individual-level health data. Research
ethics approval was therefore not required. Any future application using
person-level or M\={a}ori data requires the relevant ethics, data-governance, and
M\={a}ori-governance approvals before analysis.

\subsection*{Author contributions}
Dylan A. Mordaunt: conceptualization, methodology, software, validation,
formal analysis, investigation, data curation, visualization, writing
(original draft, review and editing), project administration, and
maintenance.

\subsection*{Use of artificial intelligence tools}
Generative AI tools were used to assist software development, manuscript
editing, consistency checking, and LaTeX preparation under author direction.
The author reviewed the resulting text, calculations, citations, and code and
retains responsibility for the work. AuthenText was used as an editorial
self-review aid and Sourceright as a provenance workflow; neither tool was
treated as evidence of factual correctness.

\subsection*{Availability of data and material}
The synthetic inputs, source code, and generated demonstration materials are available in the \pkg{vop\_poc\_nz} repository. No participant data were used.
The full protocol for this study is registered on the Open Science Framework (OSF): \url{https://osf.io/unq76}.

\subsection{Software availability}
The \pkg{vop\_poc\_nz} package is open-source and available under the Apache 2.0 license.
\begin{itemize}
    \item \textbf{PyPI}: \url{https://pypi.org/project/vop-poc-nz/}
    \item \textbf{TestPyPI}: \url{https://test.pypi.org/project/vop-poc-nz/}
    \item \textbf{Zenodo}: \url{https://zenodo.org/record/17759624}
    \item \textbf{GitHub}: \url{https://github.com/edithatogo/vop_poc_nz}
\end{itemize}

\ifdefined\JSSMODE
\else
  \bibliographystyle{plainnat}
\fi
\bibliography{ref}

\end{document}